Role of CaO addition in the local order around Erbium in SiO2-GeO2-P2O5 fiber preforms

F. d'Acapito * C. Maurizio
CNR-INFM-OGG, c/o ESRF 6, Rue Jules Horowitz, F-38043 Grenoble, France

M.C. Paul
Fibre Optics Laboratory, CGCRI 196, Raja S.C. Mullick Road, Jadavpur, Calcutta, India

Th. S. Lee W. Blanc B. Dussardier
LPMC, Universite de Nice Sophia-Antipolis, CNRS UMR6622, 28 Avenue Joseph Vallot 061008 Nice, France

author. Tel.: +33 47688 2426; fax: +33 47688 2743. E-mail address: dacapito@esrf.fr (F. d'Acapito)
* Corresponding

Abstract
The development of materials for optical signal processing represents a major issue in present technology. In this contribution we present a study on Er-doped fiber preforms where particular attention is devoted on how the addition of CaO in the glass modifies the local environment of the rare earth. The results from photoluminescence and Extended X-ray Absorption Fine Structure (EXAFS) are compared and a clear link between the width of the emission line at 1.5 μm and the amorphous/crystalline local structure around the $Er^{3+}$ ion is evidenced.

1. Introduction
Knowledge of the environment of rare earth ions in glasses is essential for the understanding of the spectral properties of these glasses and the design of new glass compositions for photonics applications. Among the rare earths, erbium is especially interesting because the $4I_{13/2} - 4I_{15/2}$ transition at ≈1.54 μm coincides with the lowest attenuation window of silica glass fiber; hence, erbium doped glasses can be used to amplify attenuated signals in optical fiber telecommunication systems. While Erbium-Doped Fiber Amplifier (EDFA) was developed 20 years ago, extensive studies are still carried out to improve its properties. In particular, linear dimensions should be reduced although the severe limitations imposed by the poor rare earth solubility into silica matrix. Moreover, spectral bandwidth should be increased to improve the Wavelength Division Multiplexing (WDM) applications. To solve these problems, various approaches have been proposed namely Er-Tm co-doping [1] or Er incorporation in nanoparticles (NP). However, to be compatible with optical waveguide applications, Rayleigh scattering must be kept at a minimum. For this purpose, Tick proposed that the particles size must be less than 15 nm with a narrow distribution size and high density [2]. There are reports on thin films containing semiconductor and metal nanoparticles acting as sensitizers, absorbing the incident light with a high crosssection and exciting erbium co-doped ions in the nanoparticle vicinity through energy transfer [3-8]. Here we propose for the first time to incorporate Er into oxide nanoparticles in optical fibers to improve the spectroscopic properties of erbium ions. To our knowledge only few studies on nano-structured silica fibers were dedicated to metal ions properties [9]. In our samples, oxide nano-structures are prepared by adjusting the composition of usual modifiers in the core of the fiber. A glass with an immiscibility gap is obtained and two phases are then formed with high and low silica content. The nanoparticles are preserved when the perform is drawn into a fiber.
An effective method to study the incorporation of Er in a matrix is by Extended X-ray Absorption Fine Structure (EXAFS) [10] as this technique provides quantitative information on the geometry and chemical composition of the environment of the element under study. The Er site has been successfully studied by this technique in a variety of matrices like crystalline Si [11,12] amorphous Si [13], Si nanoparticles [14], and silicate glass [15-17]. In the present study we have used EXAFS to describe the site of Er and to establish the crystalline/amorphous nature of its environment.

## 2. Results
### 2.1. Preform preparation and characterization

Fiber preforms were prepared by Modified Chemical Vapor Deposition (MCVD) associated to solution doping. Samples of composition $SiO_2$-$GeO_2$-$P_2O_5$-$Er_2O_3$ with and without CaO were investigated to evidence the role of Ca in the incorporation site for Er in the glass. The multiplicity of RE sites in the host matrix are known to enhance the inhomogeneous broadening of the emission/absorption lines [18] and P-codoped silicates are reported to bind in an ordered way to $Er^{3+}$ ions [19]. When preforms were drawn into fibers, NP were not directly evidenced into fibers through electron microscopy analysis; however, high attenuation losses were measured, presumably due to Rayleigh scattering induced by NP. Further analyses on this point are in progress. The preforms were cut in disks of 5 mm in diameter where the Er-doped zone is limited to a central core of 1 mm in diameter. Two samples were prepared with an $Er_2O_3$ content of 0.1 mol% and ≈1 mol% of $P_2O_5$. Sample A has 0.2 mol% of CaO whereas sample B contains no CaO. Slices of samples A and B were cut from preforms (thickness 1 mm), then successively thinned by mechanical means down to ≈10 μm and by ionic beam down to few tens of nanometers. They were characterized by electron microscopies such as Transmission Electron Microscopy (TEM) and Energy Dispersive X-Ray Micro-analysis (EDX) to analyze the chemical contrast and map the chemical composition. Fig. 1 shows a TEM image of the sample A, where a polydisperse assembly of spherical nanoparticles with a mean diameter of 50 nm can be seen. Smaller particles of 10 nm are visible and the size of the biggest ones was ≈200 nm (not shown). The formation of these particles is linked to the addition of CaO, as presented elsewhere [20]. High content of modifiers elements ([Ca] and [P] ≈30%) are present in these nano-structures. Moreover, erbium ions are found to be located only inside the nanoparticles. Spectroscopic characterizations at room temperature on the emission line associated to the $4I_{13/2} \rightarrow 4I_{15/2}$ transition at 1.54 μm were made on these samples at LPCML (Lyon, France) [21]. The results are shown in Fig. 2 where we evidence the fact that the emission spectrum of samples A is broader than that of sample B.

### 2.2. EXAFS measurements

EXAFS measurements at the Er-LIII edge (E = 8358 eV) were carried out at the GILDA-CRG beamline [22] at the European Synchrotron Radiation Facility. The monochromator was equipped with a pair of Si(3 1 1) crystals and was run in dynamical focusing mode [23]. The rejection of higher order harmonics was achieved by using a pair of Pd-coated mirrors with a cutoff energy of 20 keV. The second mirror was also used as focusing element in the vertical direction and the setup was optimized to obtain the minimum spot size on the sample; in this way 130 μm × 90 μm Full Width at Half Maximum (FWHM) were obtained. The absorption coefficient was measured in fluorescence mode by selecting the Er-Lα emission line using an high purity Ge detector. The position of each sample respect to the X-ray beam was optimized by carrying out horizontal and vertical scans and the Er doped zone resulted to be 500 μm FWHM wide. Spectra were collected at room temperature. EXAFS data are shown in Fig. 3 whereas the related Fourier Transforms are shown in Fig. 4. Sample A exhibits a single frequency signal whereas sample B presents higher frequency oscillations that are in good agreement with those observed in the spectrum of $ErPO_4$. Data were analyzed using the GNXAS [24,25] and Feff [26,27] codes. The theoretical paths were calculated using the muffin-tin approximation for the scattering potential and the Hedin-Lundquist form for its energy-dependent part. Atomic clusters were derived from the $Er_2Si_2O_7$ [28] and $ErPO_4$ [29] crystal structures. Sample A was quantitatively analyzed by using a model consisting in an $Er^{3+}$ ion linked to $SiO_4$ tetrahedra as described in [12,17]. Here we note that, due to the similar backscattering amplitude and phase of Si and P it is not possible to distinguish these two atoms in the second coordination shell. Sample B was reproduced with the $ErPO_4$ structure up to the 5th shell. The results are collected in Table 1.

## 3. Discussion

In sample A the rare earth is linked to O atoms in the first coordination shell and to Si(P) atoms in the second shell in a way similar to that already observed in silicate glasses [17,30], phosphate glasses [31,32] and in the oxide formed at purpose in crystalline Si [12]. Also the structural parameters (about seven O atoms at 2.26 Å and Si(P) atoms at 3.6 Å corresponding to a Er-O-Si(P) bond angle of ≈140°) are in good agreement with the cited literature. Si(P) atoms are visible as they belong to the same Si(P)O4 tetrahedron as the first shell O atoms but no further coordination shells are detected. This permits to state that an amorphous environment is realized around Er. On the other hand sample B presents a completely different EXAFS signal that is well comparable with the spectrum of ErPO4. This means that Er in this case is inserted in a locally well ordered phase of about a few coordination shells (around 4-5 A around the absorber). The fact that TEM on this sample reveals a uniform sample is not in contradiction with this result; it just means that this phase is not spatially extended to form nm-sized nanoparticles (in our TEM analyses the spatial resolution limited to few nm) but the ordering is extremely local, i.e. it is limited to only a few shells around the Rare Earth ion. From these considerations, the broadening of the emission spectrum observed in sample A can be attributed to an inhomogeneous broadening due to erbium located in a more disordered environment compare to sample B. Here we see that the cumulated effects of Ca and P within the Er-doped nanoparticles both amorphize the material structure around Er3+ ions and increase the fluorescence inhomogeneous broadening. Further investigations will be necessary to understand the combined role of these two ions in the amorphization of the local structure.

4. Conclusion

In this contribution we have investigated the structure around Er3+ ions in fiber preforms of composition SiO2 -GeO2-P2O5-Er2O3 with and without CaO. The addition of this component to the glass leads to the formation of calco-phospho-silicate nanoparticles. In the case without CaO the environment of Er is locally crystalline and reproduces well that observed in ErPO4 . Upon addition of Ca, amorphous nanoparticles are formed. While they contain Phosphorous, the presence of Calcium leads to an amorphization of the Erbium environment, like that observed in silicate or phosphate glasses. Correspondingly, the width of the luminescence line at 1.54 μm increases due to the inhomogeneous broadening introduced by the amorphous environment. This feature is particularly interesting in the production of materials for WDM applications.

5. Acknowledgements
This work was realized in the frame of a P2R project (Programme de Recherche en Réseau, CNRS (France) and DST (India)). We acknowledge the technical assistance of Fabrizio La Manna and Fabio d'Anca on the GILDA beamline; and Bernard Jacquier, Anne-Marie Jurdyc and Romain Peretti from LPCML (Lyon, France) for the spectroscopic measurements. Post-doctoral positions of M.C. Paul and Th. Lee Sebastian were funded by DST (Indian Department of Science and Technology, Boycast fellowship) and the MENRT (French Ministry of Research), respectively. GILDA is a project jointly financed by CNR and INFN institutes.

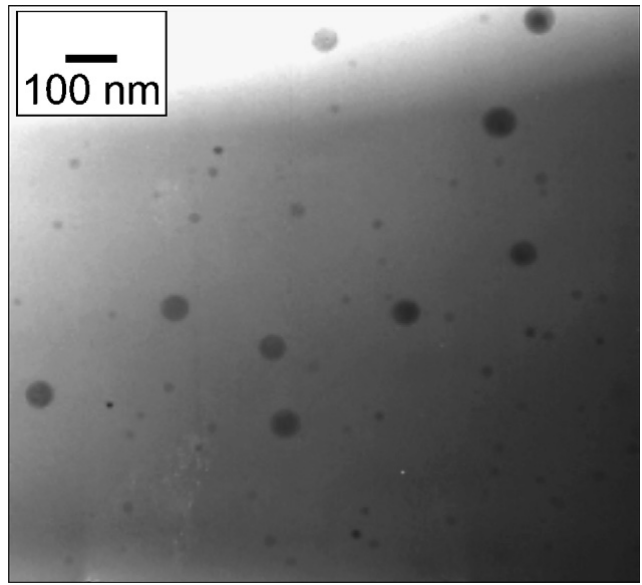

Fig. 1. TEM image from sample A (with Ca). The average particle size is 50 nm. Smaller (10 nm) particles are visible. Few bigger (≈200 nm) particles were observed, not shown on this image.

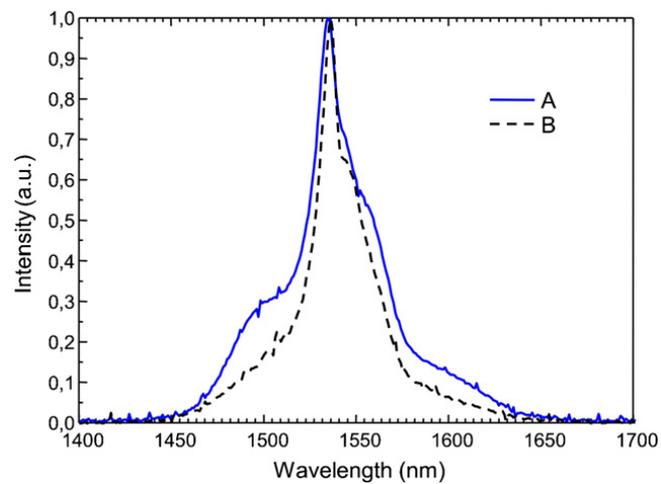

Fig. 2. Room temperature optical luminescence spectra from samples A (with Ca, continuous line) and B (without Ca, dashed line), respectively. Samples were excited at 980 nm with a laser diode the system spectral resolution was better than 1 nm. The fluorescence was measured at right angle from excitation beam.

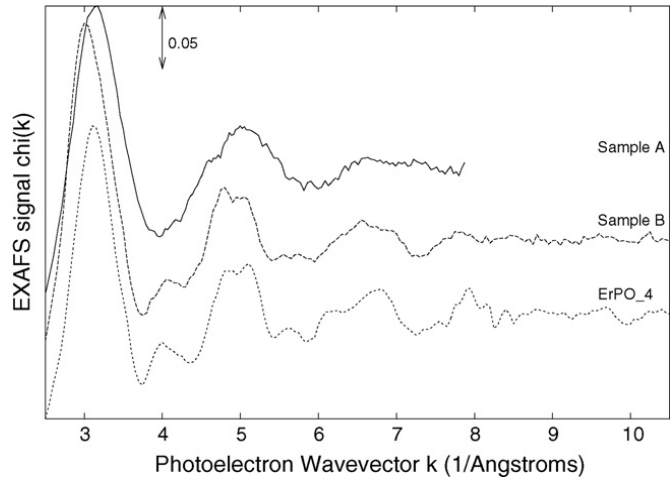

Fig. 3. EXAFS spectra of the samples.

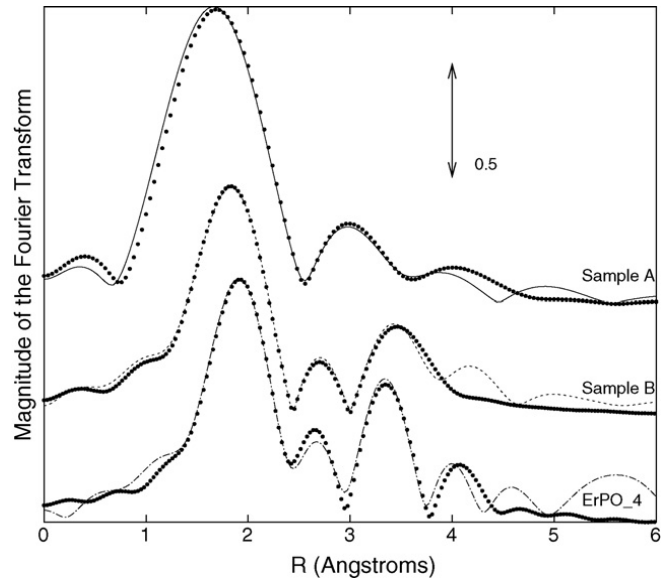

Fig. 4. Fourier Transforms of the EXAFS spectra. Transforms were carried out in the interval k ⊆ (2.5,. . ., 8.0) Å$^{-1}$ with a k$^2$ weight. The best fit curves are presented in dots.

Table 1 Results of the quantitative EXAFS analysis: bond lengths R and Debye-Waller factors σ 2 for the first three shells around Er

| Sample | $R_a$ (Å) | $\sigma_a^2$ (Å$^2$) | $R_b$ (Å) | $\sigma_b^2$ (Å$^2$) | $R_c$ (Å) | $\sigma_c^2$ (Å$^2$) |
|---|---|---|---|---|---|---|
| A | 2.23(1) | 0.012(1) | 3.60(3) | 0.015(5) | | |
| B | 2.30(2) 2.37(2) | 0.010(6) 0.010(6) | 3.01(4) 3.89(4) | 0.009(6) 0.018(6) | 3.89(4) | 0.018(6) |
| ErPO$_4$ | 2.31(2) 2.38(2) | 0.006(4) 0.006(4) | 3.03(4) 3.81(4) | 0.005(6) 0.002(6) | 3.81(4) | 0.002(4) |

The suffix a is relative to O whereas b is phosphorous in the case of samples B and ErPO$_4$ and it could be either P or Si for sample A; suffix c is Er. The number of first and second neighbors for sample A is 6.5 ± 0.2 whereas it is fixed to the crystallographic values of ErPO$_4$ (N1 = 4O, N2 = 4O, N3 = 2P, N4 = 4Er, N5 = 4P) in the other samples. Errors on the last digit are indicated in brackets.